\documentclass[sigconf]{acmart}
\AtBeginDocument{%
  }

\setcopyright{acmlicensed}
\copyrightyear{2026}
\acmYear{2026}
\acmDOI{XXXXXXX.XXXXXXX}
\acmConference[CHI '26]{CHI Conference on Human Factors in Computing Systems}{April 13--17,
  2026}{Barcelona, Spain}
\acmISBN{978-1-4503-XXXX-X/2018/06}

\begin{document}

\title[Qualitative Coding Analysis through Open-Source LLMs: A User Study and Design Recommendations]{Qualitative Coding Analysis through Open-Source Large Language Models: A User Study and Design Recommendations}

\author{Tung T. Ngo}
\authornotemark[1]
\email{thanhtung.ngo@tudublin.ie}
\orcid{0009-0004-0065-8600}
\affiliation{%
  \institution{Technological University Dublin}
  \city{Dublin}
  \country{Ireland}
}

\author{Dai Nguyen Van}

\email{dainv@neu.edu.vn}
\affiliation{%
  \institution{National Economics University}
  \city{Hanoi}
  \country{Vietnam}
}

\author{Anh-Minh Nguyen}
\email{minh.a.nguyen@ucdconnect.ie}
\orcid{0009-0005-5091-7401}
\affiliation{%
  \institution{University College Dublin}
  \city{Dublin}
  \country{Ireland}
}

\author{Phuong-Anh Do}
\email{anhdp@ftu.edu.vn}
\orcid{0009-0003-1958-881X}
\affiliation{%
 \institution{Foreign Trade University}
 \city{Hanoi}
 \country{Vietnam}
}

\author{Anh Nguyen-Quoc}
\email{anh.nguyenquoc@ucd.ie}
\orcid{0000-0001-7156-5832}
\affiliation{%
  \institution{University College Dublin}
  \city{Dublin}
  \country{Ireland}
  }

\renewcommand{\shortauthors}{Tung T. Ngo et al.}

\begin{abstract}
  Qualitative data analysis is labor-intensive, yet the privacy risks associated with commercial Large Language Models (LLMs) often preclude their use in sensitive research. To address this, we introduce ChatQDA, an on-device framework powered by open-source LLMs designed for privacy-preserving open coding. Our mixed-methods user study reveals that while participants rated the system highly for usability and perceived efficiency, they exhibited "conditional trust", valuing the tool for surface-level extraction while questioning its interpretive nuance and consistency. Furthermore, despite the technical security of local deployment, participants reported epistemic uncertainty regarding data protection, suggesting that invisible security measures are insufficient to foster trust. We conclude with design recommendations for local-first analysis tools that prioritize verifiable privacy and methodological rigor.
\end{abstract}




\maketitle

\section{Introduction}



Qualitative research is central to the social sciences because it enables scholars to explain social phenomena and build theory by surfacing meaning, process, and lived experience that are difficult to capture with predefined measures. 
Yet the craft of qualitative data analysis (QDA), particularly manual open coding, remains labour-intensive and cognitively demanding, constraining how much data researchers can analyse rigorously and transparently \cite{barros_large_2025}. Recent advances in third-party large language models (LLMs) (e.g., ChatGPT, Gemini, and Perplexity), therefore appear promising and growing in popularity: by automating aspects of qualitative analysis, they can reduce the time burden of open coding and expand analytic capacity \cite{xiao_supporting_2023}.

This promise, however, is tempered by several ethical and governance challenges. Using commercial LLMs often entails transferring confidential participant data to external servers, raising concerns about data sovereignty, re-identification risk, and compliance with institutional review board (IRB) requirements \cite{yao_survey_2024, das_security_2025}. These risks are compounded by the opacity of proprietary model architectures, including the possibility that sensitive inputs may be inadvertently memorised and later reproduced, a vulnerability documented in large-scale generative models \cite{274574}. In many qualitative research studies, especially those involving vulnerable populations or sensitive settings, these constraints might make it difficult, and sometimes impossible, to justify the use of third-party LLMs.

Open-source LLMs offer a theoretically compelling alternative for sensitive qualitative analysis. Because they can be deployed locally or within institution-controlled environments, they enable researchers to analyse qualitative data without transmitting it beyond secure boundaries, thereby strengthening privacy protection and regulatory compliance \cite{yao_survey_2024}. Their transparency and modifiability also allow adaptation to domain language and analytic approaches, including lightweight fine-tuning through Low-Rank Adaptation (LoRA) \cite{hu_lora_2022} and grounding outputs in a researcher’s corpus via Retrieval-Augmented Generation (RAG), which can reduce hallucinations and improve context retention \cite{lewis_retrieval-augmented_nodate}.

Despite their advantages, open-source LLMs can pose substantial usability and maintenance challenges for qualitative researchers, particularly those without technical expertise.
Effective use requires competence in machine learning workflows, command-line tools, and model configuration (e.g., quantization and inference parameters), creating barriers for researchers trained in interpretive or human-centered methodologies rather than computational methods \cite{barros_large_2025, lazer_computational_2020}.
Furthermore, the open-source ecosystem lacks integrated, well-designed workflows that support core qualitative research practices, forcing researchers to assemble ad hoc solutions that can undermine rigour and analytic quality \cite{xiao_supporting_2023}.
This tension, high ethical suitability but limited usability and evidence-based performance, motivates the present study.

The contributions of our research:
\begin{itemize}
    \item We developed ChatQDA, an on-device platform powered by open-source LLMs, designed to preserve data privacy while supporting qualitative researchers through an open-coding workflow.
    \item We reported findings from a user study examining how users experience and evaluate collaboration with ChatQDA during open coding, focusing on perceived usefulness, security and privacy, ease of use, trustworthiness, and behavioural intention.
    \item We distilled empirically grounded design recommendations for future open-source, LLM-based qualitative analysis tools.
    
\end{itemize}

\section{Related Work}

The integration of AI into QDA has evolved from clerical automation to interactive partnership. \cite{rietz_cody_2021} illustrated this with Cody, a system prioritizing interactive rule refinement and reflection over processing speed. Addressing collaboration, \cite{gao_coaicoder_2024} developed CoAIcoder to mediate team interpretation, though they identify a trade-off between improved consensus and reduced code diversity. Contextualizing this shift, \cite{feuston_putting_2021} argues that AI should assist rather than automate, enabling researchers to manage scale and abstraction without sacrificing methodological nuance.


The emergence of Large Language Models (LLMs) has further transformed AI into an interpretive partner capable of thematic synthesis and reflexive assistance. \cite{hamilton_exploring_2023} and \cite{morgan_exploring_2023} observe that while AI excels at identifying concrete, descriptive themes, it often misses the subtle nuances detected by human contextual knowledge; notably, Morgan found that LLMs reverse the traditional inductive cycle by presenting broad concepts first for researchers to query rather than moving from raw data to themes. Modern frameworks such as GAITA \cite{nguyen-trung_chatgpt_2025} and GATA \cite{anakok_leveraging_2025} formalize this as a guided partnership where the human remains the intellectual leader while delegating structured tasks like template refinement and data familiarization to the AI. Most recently, \cite{sinha_role_2024}  and \cite{costa_ai_2025} have advocated for AI as a reflexive co-researcher or epistemic agent. While \cite{sinha_role_2024} integrated GPT-4 to generate micro-analytic memos that surface missed data segments in Grounded Theory analysis, \cite{costa_ai_2025}’s AbductivAI model utilizes Chain-of-Thought (CoT) prompting to involve AI agents and human coders in iterative validation loops, fundamentally converting AI from a passive tool into an active intellectual contributor in distributed cognition networks.


Recent open-source advancements have significantly narrowed the performance gap with proprietary systems. Llama 3.1 \cite{sam_llama_2024} rivals GPT-4 in general reasoning, while Kimi K2 \cite{team_kimi_2025} advances agentic intelligence through autonomous tool orchestration. Innovations in reasoning-focused architectures, such as DeepSeek-R1’s \cite{guo_deepseek-r1_2025} pure reinforcement learning and the hybrid "thinking modes" of Qwen3 \cite{yang_qwen3_2025} and GLM-4.5 \cite{team_glm-45_2025}, enable self-reflection and adaptive computational allocation. These shifts are complemented by models like gpt-oss-120b \cite{openai_gpt-oss-120b_2025}, which provide customizable chain-of-thought capabilities essential for tailored research workflows.

\section{Methods}

\subsection{ChatQDA - Chatbot Powered by Open-Source LLMs}

A chat-based interface was selected for its intuitive alignment with participants' familiarity with tools like ChatGPT and Gemini, as well as its flexibility for integrating advanced LLM capabilities. 
We developed ChatQDA based on Hugging Face’s Chat UI\footnote{https://github.com/huggingface/chat-ui} framework, and the system design adheres to recommendations outlined in \cite{yan_human-ai_2024}. 
To ensure transparent reasoning, only LLMs capable of generating interpretable explanations were incorporated. 
Users can upload document files and query the system about their content, with contextual understanding enhanced by a large context window (e.g., extended token limits). 
The chat history for each session is dynamically embedded in the prompt to contextualize responses, enabling the system to refine answers iteratively based on user feedback.
We selected the gpt-oss-20b \cite{openai_gpt-oss-120b_2025} model for its efficient balance of high-tier reasoning and a compact hardware footprint optimized for on-device use.
This 20.9B-parameter MoE model utilizes MXFP4 quantization to achieve a 12.8 GiB checkpoint, allowing deployment on hardware with only 16GB of RAM.
While activating just 3.61B parameters per pass for low latency, it remains competitive on benchmarks, scoring 98.7\% on AIME 2025 (with tools) and 85.3\% on MMLU.


\subsection{User Study}

To understand how qualitative researchers experience open-source LLM assistance in open coding, we conducted a mixed-method user study. The study examines researchers’ experiences when conducting open coding manually and when collaborating with \textbf{ChatQDA} during the same analytic task.

\subsubsection{Participants}

We deliberately recruited two user groups as a comparative case study \cite{1970586434844156305}. The first group comprised novice qualitative researchers (fewer than two years of experience with qualitative data analysis and limited prior use of AI for QDA). The second group comprised more experienced qualitative researchers (at least two years of qualitative research experience and some prior exposure to AI-supported QDA). In total, we conducted four sessions (two participants per group).
Before participation, we explained the study purpose and procedures and obtained informed consent. To reduce the risk that participants’ feedback would be driven primarily by first impressions of the interface (as indicated in our pilot), we provided a detailed guide describing ChatQDA’s functions, usage scenarios and allowed participants to try the system several times prior to the study task.

\subsubsection{User Task and Requirements}

Each participant conducted two open-coding activities using the same dataset of a 5-page semi-structured interview transcript.

\begin{itemize}
    \item Task 1 - Manual open coding (baseline): Participants conducted open coding, following their usual procedures. This stage served both to familiarise participants with the dataset and to establish a baseline for comparison.
    \item Task 2 - ChatQDA-assisted open coding. Participants then collaborated with ChatQDA to open-code the same materials for 1 additional hour. 
\end{itemize}

\subsubsection{Data Collection}

Following the two coding activities, participants completed a short post-task survey administered in English via Google Forms. The survey used multi-item, 5-point Likert scales (1 = strongly disagree; 5 = strongly agree) to measure perceived security and privacy \cite{carlos_roca_importance_2009}, perceived ease of use, perceived usefulness \cite{davis_perceived_1989}, and trust \cite{choon_ling_perceived_2011}. 
We then conducted a semi-structured interview with each participant to probe the reasoning behind their survey responses and to elicit concrete examples from the session. Interviews were conducted in person, lasted approximately 40 minutes, audio-recorded, transcribed using OpenAI Whisper \cite{10.5555/3618408.3619590}, and manually checked for accuracy.

\subsubsection{Data Analysis}

Survey responses were analysed using descriptive statistics to summarise patterns across constructs. Interview transcripts and ChatQDA interaction logs were analysed using deductive thematic analysis [3], guided by the survey constructs (perceived usefulness, ease of use, security and privacy, and trustworthiness). We began with participant-level coding to develop initial codes and construct-level themes, then iteratively refined the coding scheme and theme definitions through discussion among the co-authors. The resulting themes characterise opportunities and challenges in LLM-supported open coding and inform design recommendations for future qualitative analysis tools.

\section{Key Findings}

From the study, we were able to gather two kinds of data: quantitative data from the surveys and qualitative data from the think-aloud sessions and interviews, which are described in detail below for each major theme.

\subsection{Perceived Security}


Perceived security ratings clustered around the neutral midpoint ($Median=3$), reflecting cautious confidence with no strong endorsement. Items regarding preventing third-party modification and interception got slightly higher ($M=3.0$) than those concerning general security measures and unauthorized access ($M=2.75$). 
Overall, participants did not express strong security concerns, but neither did they strongly endorse the system’s security, suggesting that perceived security may depend on clearer, more concrete assurance (e.g., transparent explanations of safeguards, evidence of local-only processing, and auditable protections).


The largely moderate (neutral) security ratings appear to reflect epistemic uncertainty rather than a negative judgment. One participant explicitly framed their midpoint choice because of limited technical expertise: \textit{“Although I’m a qualitative researcher, I’m not a technology engineer… I don’t really trust that this platform can prevent a third party from modifying the data I process. I’m not sure.”}

\subsection{Perceived Privacy}

Across the four perceived privacy items, responses clustered in the upper-mid range (3–4), indicating moderate to elevated privacy concern when using the ChatQDA. Concern was highest for data collection and privacy during use (75\% agree; $M = 3.75$), followed by potential unauthorised secondary use of research data ($M = 3.50$). Concern was lowest for unauthorised sharing with other entities ($M = 3.25$).
Overall, participants’ privacy concerns centred more on how much data the system collects and how data are handled during processing, rather than on explicit third-party sharing.


Beyond limited technical expertise, a plausible explanation is the gap between the theoretical privacy benefits and the practical realities of open-source LLM use. While participants recognised that local deployment should, in principle, strengthen privacy protections, they remained cautious because they could not independently verify how models are built, distributed, and maintained, or how data might be handled in practice. As one participant noted: \textit{“In theory, open-source LLMs may protect user data better than closed ones, but in practice we can’t know how developers use the information, or whether an open-source model contains malicious code to take user data.”}

\subsection{Perceived Ease of Use}

Across the perceived ease-of-use items, responses were generally positive, indicating that ChatQDA was easy to learn and understand. 
All participants rated 4 (agree) for “Learning to use this technology for qualitative analysis is easy for me” and for “My interaction with this technology for qualitative analysis is clear and understandable”
Perceived ease of becoming skillful was more mixed (25\% = 2, 50\% = 3, 25\% = 4; M = 3.00), suggesting strong initial learnability but some uncertainty about developing higher proficiency.



The qualitative comments help explain why participants rated ChatQDA as easy to learn and clear to interact with. A novice researcher emphasised prompting as an accessible skill: they found it straightforward to generate and iterate multiple prompts, experimenting with keywords and instructions to progressively improve the output.

\subsection{Perceived Usefulness}

Across perceived usefulness items, responses were generally positive (mostly 3–5), indicating that ChatQDA was viewed as potentially helpful for qualitative analysis, particularly for efficiency. Ratings for overall usefulness clustered around neutral to agree ($M = 3.00$), and were slightly higher for making qualitative analysis easier ($M = 3.25$). The strongest endorsement was accomplishing tasks more quickly (50\% rated 5; $M = 3.75, Md = 4$), while perceived performance improvement showed greater variability (evenly distributed across 2–5; $M = 3.50$).


Overall, data from all sources show that participants anticipated clear time-saving benefits, while views on broader performance gains were more varied, suggesting that usefulness may be most immediate for efficiency and may depend on longer-term use or stronger evidence of improvements in analytic quality.

\subsection{Trustworthiness}

Across trustworthiness items, ratings clustered around the midpoint, indicating moderate trust in ChatQDA. Overall trust was neutral to slightly positive ($M = 3.00$), while perceived reputation was lowest ($M = 2.75$), suggesting ChatQDA was not yet seen as an established research tool. Perceived competence/effectiveness was more favourable ($M = 3.25, Md = 3.5$). Reliability ratings were mixed, with half of the participants agreeing but one participant selected 1 ($M = 3.00, Md = 3.5$).

Qualitative data help explain this pattern of conditional trust. Participants emphasised that trustworthiness varied by sub-task: excerpt extraction was seen as acceptable, whereas labelling and first-order coding were viewed as less dependable. 
As one participant explained: \textit{“It depends on the task. For excerpt extraction, the quantitative accuracy is about 50–60\%, but in terms of quality, the researcher still needs to review the data, because open-source models often only identify surface-level, direct statements that are close together in the conversation. If the interviewee speaks indirectly, stating a claim in the first sentence but giving an illustrative example at the end, the AI often captures only the first sentence and ignores the example.”} 
Trust was further weakened by perceived hallucination and inconsistencies across analytic steps, illustrated by one participant’s observation: \textit{“For label extraction, the AI tends to perform much worse and shows signs of hallucination. For example, it identifies only 8 excerpts, but when assigning first-order codes it becomes 10. Usually it assigns one code per excerpt; the AI cannot identify meaning-based relationships across excerpts to group them under the same code.”} 
Participants also highlighted sensitivity to prompts as a threat to consistency: \textit{“Consistency is only slightly above average. When I change the instructions, the results change immediately, and that can push the outputs further apart. For a less experienced qualitative researcher, this could be confusing when looking at the results produced.”} 
Finally, while the system was perceived as effective for surface-level meaning \textit{“When using this platform, I feel that at the direct, surface level of meaning it processes quite quickly and is relatively effective”} participants remained sceptical about its ability to support deeper interpretive work: \textit{“I feel that for deeper layers of meaning, implicit meanings and meanings that require a broader context, reading across more parts of the data, and higher inference, the platform is not very good yet.” } Together, these insights suggest that participants’ trust in ChatQDA was bounded: the tool was valued as a fast assistant for explicit content, but viewed as less reliable for nuanced interpretation and consistent coding without careful researcher oversight.

\section{Discussions}

Overall, participants found ChatQDA easy to use and potentially useful, especially for speeding up open-coding tasks. However, they expressed cautious confidence about security, moderate privacy concern, and conditional trust in the system’s outputs, particularly for deeper interpretation and consistent coding. These tensions help explain why behavioural intention was generally positive but polarised (most willing to adopt, one clear outlier unwilling to use/recommend).

\subsection{Make Privacy and Security Visible}

Even with on-device deployment, participants did not strongly endorse security or privacy. Their ratings reflect uncertainty and limited ability to verify what the system does with data.
Improve ChatQDA by: (1) adding clear in-app statements and controls on what is stored, logged, and how to delete it; (2) showing a “local-only/no network calls” indicator; and (3) providing model provenance (version/source/checksum) to reduce concerns about open-source supply-chain risks.

\subsection{Build Trust through Traceability and Consistency}

Trustworthiness varied by sub-task: excerpt extraction was acceptable, but labelling/first-order coding raised concerns about hallucination, inconsistency across steps, and sensitivity to prompt changes. Participants also noted the system handled surface meaning better than latent meaning requiring broader context. Improve ChatQDA by: (1) forcing evidence-based outputs (highlighted spans/line references for every excerpt/code); (2) adding guardrails to prevent “new” excerpts/codes appearing across stages; (3) enabling reproducible runs (saved prompts/settings) and a lockable codebook; and (4) supporting grouping related excerpts under shared codes.

\subsection{Move from "Fast" to "Methodologically Defensible"}

Usefulness was strongest for efficiency, while perceived performance gains were more varied, suggesting ChatQDA’s value is clearest as a time-saving assistant, but broader adoption depends on supporting analytic rigour. Improve ChatQDA by: integrating memoing and audit trails, linking codes to evidence, and providing lightweight prompt templates for common qualitative tasks to help users become skilful (not just competent).

\begin{acks}
To Robert, for the bagels and explaining CMYK and color spaces.
\end{acks}

\bibliographystyle{ACM-Reference-Format}
\bibliography{CHI}


\begin{thebibliography}{29}


\ifx \showCODEN    \undefined \def \showCODEN     #1{\unskip}     \fi
\ifx \showISBNx    \undefined \def \showISBNx     #1{\unskip}     \fi
\ifx \showISBNxiii \undefined \def \showISBNxiii  #1{\unskip}     \fi
\ifx \showISSN     \undefined \def \showISSN      #1{\unskip}     \fi
\ifx \showLCCN     \undefined \def \showLCCN      #1{\unskip}     \fi
\ifx \shownote     \undefined \def \shownote      #1{#1}          \fi
\ifx \showarticletitle \undefined \def \showarticletitle #1{#1}   \fi
\ifx \showURL      \undefined \def \showURL       {\relax}        \fi
\providecommand\bibfield[2]{#2}
\providecommand\bibinfo[2]{#2}
\providecommand\natexlab[1]{#1}
\providecommand\showeprint[2][]{arXiv:#2}

\bibitem[Anakok et~al\mbox{.}(2025)]%
        {anakok_leveraging_2025}
\bibfield{author}{\bibinfo{person}{Isil Anakok}, \bibinfo{person}{Andrew Katz}, \bibinfo{person}{Kai~Jun Chew}, {and} \bibinfo{person}{Holly Matusovich}.} \bibinfo{year}{2025}\natexlab{}.
\newblock \showarticletitle{Leveraging {Generative} {Text} {Models} and {Natural} {Language} {Processing} to {Perform} {Traditional} {Thematic} {Data} {Analysis}}.
\newblock \bibinfo{journal}{\emph{International Journal of Qualitative Methods}}  \bibinfo{volume}{24} (\bibinfo{date}{April} \bibinfo{year}{2025}), \bibinfo{pages}{16094069251338898}.
\newblock
\showISSN{1609-4069, 1609-4069}
\href{https://doi.org/10.1177/16094069251338898}{doi:\nolinkurl{10.1177/16094069251338898}}


\bibitem[Barros et~al\mbox{.}(2025)]%
        {barros_large_2025}
\bibfield{author}{\bibinfo{person}{Cauã~Ferreira Barros}, \bibinfo{person}{Bruna~Borges Azevedo}, \bibinfo{person}{Valdemar~Vicente Graciano~Neto}, \bibinfo{person}{Mohamad Kassab}, \bibinfo{person}{Marcos Kalinowski}, \bibinfo{person}{Hugo Alexandre~D. Do~Nascimento}, {and} \bibinfo{person}{Michelle~C.G.S.P. Bandeira}.} \bibinfo{year}{2025}\natexlab{}.
\newblock \showarticletitle{Large {Language} {Model} for {Qualitative} {Research}: {A} {Systematic} {Mapping} {Study}}. In \bibinfo{booktitle}{\emph{2025 {IEEE}/{ACM} {International} {Workshop} on {Methodological} {Issues} with {Empirical} {Studies} in {Software} {Engineering} ({WSESE})}}. \bibinfo{publisher}{IEEE}, \bibinfo{address}{Ottawa, ON, Canada}, \bibinfo{pages}{48--55}.
\newblock
\showISBNx{979-8-3315-0225-6}
\href{https://doi.org/10.1109/WSESE66602.2025.00015}{doi:\nolinkurl{10.1109/WSESE66602.2025.00015}}


\bibitem[Carlini et~al\mbox{.}(2021)]%
        {274574}
\bibfield{author}{\bibinfo{person}{Nicholas Carlini}, \bibinfo{person}{Florian Tram{\`e}r}, \bibinfo{person}{Eric Wallace}, \bibinfo{person}{Matthew Jagielski}, \bibinfo{person}{Ariel Herbert-Voss}, \bibinfo{person}{Katherine Lee}, \bibinfo{person}{Adam Roberts}, \bibinfo{person}{Tom Brown}, \bibinfo{person}{Dawn Song}, \bibinfo{person}{{\'U}lfar Erlingsson}, \bibinfo{person}{Alina Oprea}, {and} \bibinfo{person}{Colin Raffel}.} \bibinfo{year}{2021}\natexlab{}.
\newblock \showarticletitle{Extracting Training Data from Large Language Models}. In \bibinfo{booktitle}{\emph{30th USENIX Security Symposium (USENIX Security 21)}}. \bibinfo{publisher}{USENIX Association}, \bibinfo{pages}{2633--2650}.
\newblock
\showISBNx{978-1-939133-24-3}
\urldef\tempurl%
\url{https://www.usenix.org/conference/usenixsecurity21/presentation/carlini-extracting}
\showURL{%
\tempurl}


\bibitem[Carlos~Roca et~al\mbox{.}(2009)]%
        {carlos_roca_importance_2009}
\bibfield{author}{\bibinfo{person}{Juan Carlos~Roca}, \bibinfo{person}{Juan José~García}, {and} \bibinfo{person}{Juan José de~la Vega}.} \bibinfo{year}{2009}\natexlab{}.
\newblock \showarticletitle{The importance of perceived trust, security and privacy in online trading systems}.
\newblock \bibinfo{journal}{\emph{Information Management \& Computer Security}} \bibinfo{volume}{17}, \bibinfo{number}{2} (\bibinfo{date}{June} \bibinfo{year}{2009}), \bibinfo{pages}{96--113}.
\newblock
\showISSN{0968-5227}
\href{https://doi.org/10.1108/09685220910963983}{doi:\nolinkurl{10.1108/09685220910963983}}
\newblock
\shownote{\_eprint: https://www.emerald.com/ics/article-pdf/17/2/96/1208198/09685220910963983.pdf}.


\bibitem[Choon~Ling et~al\mbox{.}(2011)]%
        {choon_ling_perceived_2011}
\bibfield{author}{\bibinfo{person}{Kwek Choon~Ling}, \bibinfo{person}{Dazmin Bin~Daud}, \bibinfo{person}{Tan Hoi~Piew}, \bibinfo{person}{Kay~Hooi Keoy}, {and} \bibinfo{person}{Padzil Hassan}.} \bibinfo{year}{2011}\natexlab{}.
\newblock \showarticletitle{Perceived {Risk}, {Perceived} {Technology}, {Online} {Trust} for the {Online} {Purchase} {Intention} in {Malaysia}}.
\newblock \bibinfo{journal}{\emph{International Journal of Business and Management}} \bibinfo{volume}{6}, \bibinfo{number}{6} (\bibinfo{date}{June} \bibinfo{year}{2011}), \bibinfo{pages}{p167}.
\newblock
\showISSN{1833-8119, 1833-3850}
\href{https://doi.org/10.5539/ijbm.v6n6p167}{doi:\nolinkurl{10.5539/ijbm.v6n6p167}}


\bibitem[Costa et~al\mbox{.}(2025)]%
        {costa_ai_2025}
\bibfield{author}{\bibinfo{person}{António~Pedro Costa}, \bibinfo{person}{Grzegorz Bryda}, \bibinfo{person}{Prokopis~A. Christou}, {and} \bibinfo{person}{Judita Kasperiuniene}.} \bibinfo{year}{2025}\natexlab{}.
\newblock \showarticletitle{{AI} as a {Co}-researcher in the {Qualitative} {Research} {Workflow}: {Transforming} {Human}-{AI} {Collaboration}}.
\newblock \bibinfo{journal}{\emph{International Journal of Qualitative Methods}}  \bibinfo{volume}{24} (\bibinfo{date}{Sept.} \bibinfo{year}{2025}), \bibinfo{pages}{16094069251383739}.
\newblock
\showISSN{1609-4069, 1609-4069}
\href{https://doi.org/10.1177/16094069251383739}{doi:\nolinkurl{10.1177/16094069251383739}}


\bibitem[Das et~al\mbox{.}(2025)]%
        {das_security_2025}
\bibfield{author}{\bibinfo{person}{Badhan~Chandra Das}, \bibinfo{person}{M.~Hadi Amini}, {and} \bibinfo{person}{Yanzhao Wu}.} \bibinfo{year}{2025}\natexlab{}.
\newblock \showarticletitle{Security and {Privacy} {Challenges} of {Large} {Language} {Models}: {A} {Survey}}.
\newblock \bibinfo{journal}{\emph{Comput. Surveys}} \bibinfo{volume}{57}, \bibinfo{number}{6} (\bibinfo{date}{June} \bibinfo{year}{2025}), \bibinfo{pages}{1--39}.
\newblock
\showISSN{0360-0300, 1557-7341}
\href{https://doi.org/10.1145/3712001}{doi:\nolinkurl{10.1145/3712001}}


\bibitem[Davis(1989)]%
        {davis_perceived_1989}
\bibfield{author}{\bibinfo{person}{Fred~D. Davis}.} \bibinfo{year}{1989}\natexlab{}.
\newblock \showarticletitle{Perceived {Usefulness}, {Perceived} {Ease} of {Use}, and {User} {Acceptance} of {Information} {Technology}}.
\newblock \bibinfo{journal}{\emph{Management Information Systems Quarterly}} \bibinfo{volume}{13}, \bibinfo{number}{3} (\bibinfo{date}{Sept.} \bibinfo{year}{1989}), \bibinfo{pages}{319--340}.
\newblock
\showISSN{0276-7783}
\href{https://doi.org/10.2307/249008}{doi:\nolinkurl{10.2307/249008}}
\newblock
\shownote{\_eprint: https://misq.umn.edu/misq/article-pdf/13/3/319/903/6\_davis.pdf}.


\bibitem[Feuston and Brubaker(2021)]%
        {feuston_putting_2021}
\bibfield{author}{\bibinfo{person}{Jessica~L. Feuston} {and} \bibinfo{person}{Jed~R. Brubaker}.} \bibinfo{year}{2021}\natexlab{}.
\newblock \showarticletitle{Putting {Tools} in {Their} {Place}: {The} {Role} of {Time} and {Perspective} in {Human}-{AI} {Collaboration} for {Qualitative} {Analysis}}.
\newblock \bibinfo{journal}{\emph{Proceedings of the ACM on Human-Computer Interaction}} \bibinfo{volume}{5}, \bibinfo{number}{CSCW2} (\bibinfo{date}{Oct.} \bibinfo{year}{2021}), \bibinfo{pages}{1--25}.
\newblock
\showISSN{2573-0142}
\href{https://doi.org/10.1145/3479856}{doi:\nolinkurl{10.1145/3479856}}


\bibitem[Gao et~al\mbox{.}(2024)]%
        {gao_coaicoder_2024}
\bibfield{author}{\bibinfo{person}{Jie Gao}, \bibinfo{person}{Kenny Tsu~Wei Choo}, \bibinfo{person}{Junming Cao}, \bibinfo{person}{Roy Ka-Wei Lee}, {and} \bibinfo{person}{Simon Perrault}.} \bibinfo{year}{2024}\natexlab{}.
\newblock \showarticletitle{{CoAIcoder}: {Examining} the {Effectiveness} of {AI}-assisted {Human}-to-{Human} {Collaboration} in {Qualitative} {Analysis}}.
\newblock \bibinfo{journal}{\emph{ACM Transactions on Computer-Human Interaction}} \bibinfo{volume}{31}, \bibinfo{number}{1} (\bibinfo{date}{Feb.} \bibinfo{year}{2024}), \bibinfo{pages}{1--38}.
\newblock
\showISSN{1073-0516, 1557-7325}
\href{https://doi.org/10.1145/3617362}{doi:\nolinkurl{10.1145/3617362}}


\bibitem[Guo et~al\mbox{.}(2025)]%
        {guo_deepseek-r1_2025}
\bibfield{author}{\bibinfo{person}{Daya Guo}, \bibinfo{person}{Dejian Yang}, \bibinfo{person}{Haowei Zhang}, \bibinfo{person}{Junxiao Song}, \bibinfo{person}{Peiyi Wang}, \bibinfo{person}{Qihao Zhu}, \bibinfo{person}{Runxin Xu}, \bibinfo{person}{Ruoyu Zhang}, \bibinfo{person}{Shirong Ma}, \bibinfo{person}{Xiao Bi}, \bibinfo{person}{Xiaokang Zhang}, \bibinfo{person}{Xingkai Yu}, \bibinfo{person}{Yu Wu}, \bibinfo{person}{Z.~F. Wu}, \bibinfo{person}{Zhibin Gou}, \bibinfo{person}{Zhihong Shao}, \bibinfo{person}{Zhuoshu Li}, \bibinfo{person}{Ziyi Gao}, \bibinfo{person}{Aixin Liu}, \bibinfo{person}{Bing Xue}, \bibinfo{person}{Bingxuan Wang}, \bibinfo{person}{Bochao Wu}, \bibinfo{person}{Bei Feng}, \bibinfo{person}{Chengda Lu}, \bibinfo{person}{Chenggang Zhao}, \bibinfo{person}{Chengqi Deng}, \bibinfo{person}{Chong Ruan}, \bibinfo{person}{Damai Dai}, \bibinfo{person}{Deli Chen}, \bibinfo{person}{Dongjie Ji}, \bibinfo{person}{Erhang Li}, \bibinfo{person}{Fangyun Lin}, \bibinfo{person}{Fucong Dai},
  \bibinfo{person}{Fuli Luo}, \bibinfo{person}{Guangbo Hao}, \bibinfo{person}{Guanting Chen}, \bibinfo{person}{Guowei Li}, \bibinfo{person}{H. Zhang}, \bibinfo{person}{Hanwei Xu}, \bibinfo{person}{Honghui Ding}, \bibinfo{person}{Huazuo Gao}, \bibinfo{person}{Hui Qu}, \bibinfo{person}{Hui Li}, \bibinfo{person}{Jianzhong Guo}, \bibinfo{person}{Jiashi Li}, \bibinfo{person}{Jingchang Chen}, \bibinfo{person}{Jingyang Yuan}, \bibinfo{person}{Jinhao Tu}, \bibinfo{person}{Junjie Qiu}, \bibinfo{person}{Junlong Li}, \bibinfo{person}{J.~L. Cai}, \bibinfo{person}{Jiaqi Ni}, \bibinfo{person}{Jian Liang}, \bibinfo{person}{Jin Chen}, \bibinfo{person}{Kai Dong}, \bibinfo{person}{Kai Hu}, \bibinfo{person}{Kaichao You}, \bibinfo{person}{Kaige Gao}, \bibinfo{person}{Kang Guan}, \bibinfo{person}{Kexin Huang}, \bibinfo{person}{Kuai Yu}, \bibinfo{person}{Lean Wang}, \bibinfo{person}{Lecong Zhang}, \bibinfo{person}{Liang Zhao}, \bibinfo{person}{Litong Wang}, \bibinfo{person}{Liyue Zhang}, \bibinfo{person}{Lei Xu},
  \bibinfo{person}{Leyi Xia}, \bibinfo{person}{Mingchuan Zhang}, \bibinfo{person}{Minghua Zhang}, \bibinfo{person}{Minghui Tang}, \bibinfo{person}{Mingxu Zhou}, \bibinfo{person}{Meng Li}, \bibinfo{person}{Miaojun Wang}, \bibinfo{person}{Mingming Li}, \bibinfo{person}{Ning Tian}, \bibinfo{person}{Panpan Huang}, \bibinfo{person}{Peng Zhang}, \bibinfo{person}{Qiancheng Wang}, \bibinfo{person}{Qinyu Chen}, \bibinfo{person}{Qiushi Du}, \bibinfo{person}{Ruiqi Ge}, \bibinfo{person}{Ruisong Zhang}, \bibinfo{person}{Ruizhe Pan}, \bibinfo{person}{Runji Wang}, \bibinfo{person}{R.~J. Chen}, \bibinfo{person}{R.~L. Jin}, \bibinfo{person}{Ruyi Chen}, \bibinfo{person}{Shanghao Lu}, \bibinfo{person}{Shangyan Zhou}, \bibinfo{person}{Shanhuang Chen}, \bibinfo{person}{Shengfeng Ye}, \bibinfo{person}{Shiyu Wang}, \bibinfo{person}{Shuiping Yu}, \bibinfo{person}{Shunfeng Zhou}, \bibinfo{person}{Shuting Pan}, \bibinfo{person}{S.~S. Li}, \bibinfo{person}{Shuang Zhou}, \bibinfo{person}{Shaoqing Wu}, \bibinfo{person}{Tao Yun},
  \bibinfo{person}{Tian Pei}, \bibinfo{person}{Tianyu Sun}, \bibinfo{person}{T. Wang}, \bibinfo{person}{Wangding Zeng}, \bibinfo{person}{Wen Liu}, \bibinfo{person}{Wenfeng Liang}, \bibinfo{person}{Wenjun Gao}, \bibinfo{person}{Wenqin Yu}, \bibinfo{person}{Wentao Zhang}, \bibinfo{person}{W.~L. Xiao}, \bibinfo{person}{Wei An}, \bibinfo{person}{Xiaodong Liu}, \bibinfo{person}{Xiaohan Wang}, \bibinfo{person}{Xiaokang Chen}, \bibinfo{person}{Xiaotao Nie}, \bibinfo{person}{Xin Cheng}, \bibinfo{person}{Xin Liu}, \bibinfo{person}{Xin Xie}, \bibinfo{person}{Xingchao Liu}, \bibinfo{person}{Xinyu Yang}, \bibinfo{person}{Xinyuan Li}, \bibinfo{person}{Xuecheng Su}, \bibinfo{person}{Xuheng Lin}, \bibinfo{person}{X.~Q. Li}, \bibinfo{person}{Xiangyue Jin}, \bibinfo{person}{Xiaojin Shen}, \bibinfo{person}{Xiaosha Chen}, \bibinfo{person}{Xiaowen Sun}, \bibinfo{person}{Xiaoxiang Wang}, \bibinfo{person}{Xinnan Song}, \bibinfo{person}{Xinyi Zhou}, \bibinfo{person}{Xianzu Wang}, \bibinfo{person}{Xinxia Shan},
  \bibinfo{person}{Y.~K. Li}, \bibinfo{person}{Y.~Q. Wang}, \bibinfo{person}{Y.~X. Wei}, \bibinfo{person}{Yang Zhang}, \bibinfo{person}{Yanhong Xu}, \bibinfo{person}{Yao Li}, \bibinfo{person}{Yao Zhao}, \bibinfo{person}{Yaofeng Sun}, \bibinfo{person}{Yaohui Wang}, \bibinfo{person}{Yi Yu}, \bibinfo{person}{Yichao Zhang}, \bibinfo{person}{Yifan Shi}, \bibinfo{person}{Yiliang Xiong}, \bibinfo{person}{Ying He}, \bibinfo{person}{Yishi Piao}, \bibinfo{person}{Yisong Wang}, \bibinfo{person}{Yixuan Tan}, \bibinfo{person}{Yiyang Ma}, \bibinfo{person}{Yiyuan Liu}, \bibinfo{person}{Yongqiang Guo}, \bibinfo{person}{Yuan Ou}, \bibinfo{person}{Yuduan Wang}, \bibinfo{person}{Yue Gong}, \bibinfo{person}{Yuheng Zou}, \bibinfo{person}{Yujia He}, \bibinfo{person}{Yunfan Xiong}, \bibinfo{person}{Yuxiang Luo}, \bibinfo{person}{Yuxiang You}, \bibinfo{person}{Yuxuan Liu}, \bibinfo{person}{Yuyang Zhou}, \bibinfo{person}{Y.~X. Zhu}, \bibinfo{person}{Yanping Huang}, \bibinfo{person}{Yaohui Li}, \bibinfo{person}{Yi Zheng},
  \bibinfo{person}{Yuchen Zhu}, \bibinfo{person}{Yunxian Ma}, \bibinfo{person}{Ying Tang}, \bibinfo{person}{Yukun Zha}, \bibinfo{person}{Yuting Yan}, \bibinfo{person}{Z.~Z. Ren}, \bibinfo{person}{Zehui Ren}, \bibinfo{person}{Zhangli Sha}, \bibinfo{person}{Zhe Fu}, \bibinfo{person}{Zhean Xu}, \bibinfo{person}{Zhenda Xie}, \bibinfo{person}{Zhengyan Zhang}, \bibinfo{person}{Zhewen Hao}, \bibinfo{person}{Zhicheng Ma}, \bibinfo{person}{Zhigang Yan}, \bibinfo{person}{Zhiyu Wu}, \bibinfo{person}{Zihui Gu}, \bibinfo{person}{Zijia Zhu}, \bibinfo{person}{Zijun Liu}, \bibinfo{person}{Zilin Li}, \bibinfo{person}{Ziwei Xie}, \bibinfo{person}{Ziyang Song}, \bibinfo{person}{Zizheng Pan}, \bibinfo{person}{Zhen Huang}, \bibinfo{person}{Zhipeng Xu}, \bibinfo{person}{Zhongyu Zhang}, {and} \bibinfo{person}{Zhen Zhang}.} \bibinfo{year}{2025}\natexlab{}.
\newblock \showarticletitle{{DeepSeek}-{R1} incentivizes reasoning in {LLMs} through reinforcement learning}.
\newblock \bibinfo{journal}{\emph{Nature}} \bibinfo{volume}{645}, \bibinfo{number}{8081} (\bibinfo{date}{Sept.} \bibinfo{year}{2025}), \bibinfo{pages}{633--638}.
\newblock
\showISSN{0028-0836, 1476-4687}
\href{https://doi.org/10.1038/s41586-025-09422-z}{doi:\nolinkurl{10.1038/s41586-025-09422-z}}


\bibitem[Hamilton et~al\mbox{.}(2023)]%
        {hamilton_exploring_2023}
\bibfield{author}{\bibinfo{person}{Leah Hamilton}, \bibinfo{person}{Desha Elliott}, \bibinfo{person}{Aaron Quick}, \bibinfo{person}{Simone Smith}, {and} \bibinfo{person}{Victoria Choplin}.} \bibinfo{year}{2023}\natexlab{}.
\newblock \showarticletitle{Exploring the {Use} of {AI} in {Qualitative} {Analysis}: {A} {Comparative} {Study} of {Guaranteed} {Income} {Data}}.
\newblock \bibinfo{journal}{\emph{International Journal of Qualitative Methods}}  \bibinfo{volume}{22} (\bibinfo{date}{Oct.} \bibinfo{year}{2023}), \bibinfo{pages}{16094069231201504}.
\newblock
\showISSN{1609-4069, 1609-4069}
\href{https://doi.org/10.1177/16094069231201504}{doi:\nolinkurl{10.1177/16094069231201504}}


\bibitem[Hu et~al\mbox{.}(2022)]%
        {hu_lora_2022}
\bibfield{author}{\bibinfo{person}{Edward Hu}, \bibinfo{person}{Yelong Shen}, \bibinfo{person}{Phillip Wallis}, \bibinfo{person}{Zeyuan Allen-Zhu}, \bibinfo{person}{Yuanzhi Li}, \bibinfo{person}{Shean Wang}, \bibinfo{person}{Lu Wang}, {and} \bibinfo{person}{Weizhu Chen}.} \bibinfo{year}{2022}\natexlab{}.
\newblock \showarticletitle{{LORA}: {LOW}-{RANK} {ADAPTATION} {OF} {LARGE} {LAN}- {GUAGE} {MODELS}}.
\newblock  (\bibinfo{year}{2022}).
\newblock


\bibitem[Lazer et~al\mbox{.}(2020)]%
        {lazer_computational_2020}
\bibfield{author}{\bibinfo{person}{David M.~J. Lazer}, \bibinfo{person}{Alex Pentland}, \bibinfo{person}{Duncan~J. Watts}, \bibinfo{person}{Sinan Aral}, \bibinfo{person}{Susan Athey}, \bibinfo{person}{Noshir Contractor}, \bibinfo{person}{Deen Freelon}, \bibinfo{person}{Sandra Gonzalez-Bailon}, \bibinfo{person}{Gary King}, \bibinfo{person}{Helen Margetts}, \bibinfo{person}{Alondra Nelson}, \bibinfo{person}{Matthew~J. Salganik}, \bibinfo{person}{Markus Strohmaier}, \bibinfo{person}{Alessandro Vespignani}, {and} \bibinfo{person}{Claudia Wagner}.} \bibinfo{year}{2020}\natexlab{}.
\newblock \showarticletitle{Computational social science: {Obstacles} and opportunities}.
\newblock \bibinfo{journal}{\emph{Science}} \bibinfo{volume}{369}, \bibinfo{number}{6507} (\bibinfo{date}{Aug.} \bibinfo{year}{2020}), \bibinfo{pages}{1060--1062}.
\newblock
\showISSN{0036-8075, 1095-9203}
\href{https://doi.org/10.1126/science.aaz8170}{doi:\nolinkurl{10.1126/science.aaz8170}}


\bibitem[Lewis et~al\mbox{.}({[n.\,d.]})]%
        {lewis_retrieval-augmented_nodate}
\bibfield{author}{\bibinfo{person}{Patrick Lewis}, \bibinfo{person}{Ethan Perez}, \bibinfo{person}{Aleksandra Piktus}, \bibinfo{person}{Fabio Petroni}, \bibinfo{person}{Vladimir Karpukhin}, \bibinfo{person}{Naman Goyal}, \bibinfo{person}{Heinrich Küttler}, \bibinfo{person}{Mike Lewis}, \bibinfo{person}{Wen-tau Yih}, \bibinfo{person}{Tim Rocktäschel}, \bibinfo{person}{Sebastian Riedel}, {and} \bibinfo{person}{Douwe Kiela}.} \bibinfo{year}{[n.\,d.]}\natexlab{}.
\newblock \showarticletitle{Retrieval-{Augmented} {Generation} for {Knowledge}-{Intensive} {NLP} {Tasks}}.
\newblock  (\bibinfo{year}{[n.\,d.]}).
\newblock


\bibitem[Morgan(2023)]%
        {morgan_exploring_2023}
\bibfield{author}{\bibinfo{person}{David~L. Morgan}.} \bibinfo{year}{2023}\natexlab{}.
\newblock \showarticletitle{Exploring the {Use} of {Artificial} {Intelligence} for {Qualitative} {Data} {Analysis}: {The} {Case} of {ChatGPT}}.
\newblock \bibinfo{journal}{\emph{International Journal of Qualitative Methods}}  \bibinfo{volume}{22} (\bibinfo{date}{Oct.} \bibinfo{year}{2023}), \bibinfo{pages}{16094069231211248}.
\newblock
\showISSN{1609-4069, 1609-4069}
\href{https://doi.org/10.1177/16094069231211248}{doi:\nolinkurl{10.1177/16094069231211248}}


\bibitem[Nguyen-Trung(2025)]%
        {nguyen-trung_chatgpt_2025}
\bibfield{author}{\bibinfo{person}{Kien Nguyen-Trung}.} \bibinfo{year}{2025}\natexlab{}.
\newblock \showarticletitle{{ChatGPT} in thematic analysis: {Can} {AI} become a research assistant in qualitative research?}
\newblock \bibinfo{journal}{\emph{Quality \& Quantity}} \bibinfo{volume}{59}, \bibinfo{number}{6} (\bibinfo{date}{Dec.} \bibinfo{year}{2025}), \bibinfo{pages}{4945--4978}.
\newblock
\showISSN{0033-5177, 1573-7845}
\href{https://doi.org/10.1007/s11135-025-02165-z}{doi:\nolinkurl{10.1007/s11135-025-02165-z}}


\bibitem[OpenAI et~al\mbox{.}(2025)]%
        {openai_gpt-oss-120b_2025}
\bibfield{author}{\bibinfo{person}{OpenAI}, \bibinfo{person}{Sandhini Agarwal}, \bibinfo{person}{Lama Ahmad}, \bibinfo{person}{Jason Ai}, \bibinfo{person}{Sam Altman}, \bibinfo{person}{Andy Applebaum}, \bibinfo{person}{Edwin Arbus}, \bibinfo{person}{Rahul~K. Arora}, \bibinfo{person}{Yu Bai}, \bibinfo{person}{Bowen Baker}, \bibinfo{person}{Haiming Bao}, \bibinfo{person}{Boaz Barak}, \bibinfo{person}{Ally Bennett}, \bibinfo{person}{Tyler Bertao}, \bibinfo{person}{Nivedita Brett}, \bibinfo{person}{Eugene Brevdo}, \bibinfo{person}{Greg Brockman}, \bibinfo{person}{Sebastien Bubeck}, \bibinfo{person}{Che Chang}, \bibinfo{person}{Kai Chen}, \bibinfo{person}{Mark Chen}, \bibinfo{person}{Enoch Cheung}, \bibinfo{person}{Aidan Clark}, \bibinfo{person}{Dan Cook}, \bibinfo{person}{Marat Dukhan}, \bibinfo{person}{Casey Dvorak}, \bibinfo{person}{Kevin Fives}, \bibinfo{person}{Vlad Fomenko}, \bibinfo{person}{Timur Garipov}, \bibinfo{person}{Kristian Georgiev}, \bibinfo{person}{Mia Glaese}, \bibinfo{person}{Tarun Gogineni},
  \bibinfo{person}{Adam Goucher}, \bibinfo{person}{Lukas Gross}, \bibinfo{person}{Katia~Gil Guzman}, \bibinfo{person}{John Hallman}, \bibinfo{person}{Jackie Hehir}, \bibinfo{person}{Johannes Heidecke}, \bibinfo{person}{Alec Helyar}, \bibinfo{person}{Haitang Hu}, \bibinfo{person}{Romain Huet}, \bibinfo{person}{Jacob Huh}, \bibinfo{person}{Saachi Jain}, \bibinfo{person}{Zach Johnson}, \bibinfo{person}{Chris Koch}, \bibinfo{person}{Irina Kofman}, \bibinfo{person}{Dominik Kundel}, \bibinfo{person}{Jason Kwon}, \bibinfo{person}{Volodymyr Kyrylov}, \bibinfo{person}{Elaine~Ya Le}, \bibinfo{person}{Guillaume Leclerc}, \bibinfo{person}{James~Park Lennon}, \bibinfo{person}{Scott Lessans}, \bibinfo{person}{Mario Lezcano-Casado}, \bibinfo{person}{Yuanzhi Li}, \bibinfo{person}{Zhuohan Li}, \bibinfo{person}{Ji Lin}, \bibinfo{person}{Jordan Liss}, \bibinfo{person}{Lily}, \bibinfo{person}{Liu}, \bibinfo{person}{Jiancheng Liu}, \bibinfo{person}{Kevin Lu}, \bibinfo{person}{Chris Lu}, \bibinfo{person}{Zoran Martinovic},
  \bibinfo{person}{Lindsay McCallum}, \bibinfo{person}{Josh McGrath}, \bibinfo{person}{Scott McKinney}, \bibinfo{person}{Aidan McLaughlin}, \bibinfo{person}{Song Mei}, \bibinfo{person}{Steve Mostovoy}, \bibinfo{person}{Tong Mu}, \bibinfo{person}{Gideon Myles}, \bibinfo{person}{Alexander Neitz}, \bibinfo{person}{Alex Nichol}, \bibinfo{person}{Jakub Pachocki}, \bibinfo{person}{Alex Paino}, \bibinfo{person}{Dana Palmie}, \bibinfo{person}{Ashley Pantuliano}, \bibinfo{person}{Giambattista Parascandolo}, \bibinfo{person}{Jongsoo Park}, \bibinfo{person}{Leher Pathak}, \bibinfo{person}{Carolina Paz}, \bibinfo{person}{Ludovic Peran}, \bibinfo{person}{Dmitry Pimenov}, \bibinfo{person}{Michelle Pokrass}, \bibinfo{person}{Elizabeth Proehl}, \bibinfo{person}{Huida Qiu}, \bibinfo{person}{Gaby Raila}, \bibinfo{person}{Filippo Raso}, \bibinfo{person}{Hongyu Ren}, \bibinfo{person}{Kimmy Richardson}, \bibinfo{person}{David Robinson}, \bibinfo{person}{Bob Rotsted}, \bibinfo{person}{Hadi Salman}, \bibinfo{person}{Suvansh
  Sanjeev}, \bibinfo{person}{Max Schwarzer}, \bibinfo{person}{D. Sculley}, \bibinfo{person}{Harshit Sikchi}, \bibinfo{person}{Kendal Simon}, \bibinfo{person}{Karan Singhal}, \bibinfo{person}{Yang Song}, \bibinfo{person}{Dane Stuckey}, \bibinfo{person}{Zhiqing Sun}, \bibinfo{person}{Philippe Tillet}, \bibinfo{person}{Sam Toizer}, \bibinfo{person}{Foivos Tsimpourlas}, \bibinfo{person}{Nikhil Vyas}, \bibinfo{person}{Eric Wallace}, \bibinfo{person}{Xin Wang}, \bibinfo{person}{Miles Wang}, \bibinfo{person}{Olivia Watkins}, \bibinfo{person}{Kevin Weil}, \bibinfo{person}{Amy Wendling}, \bibinfo{person}{Kevin Whinnery}, \bibinfo{person}{Cedric Whitney}, \bibinfo{person}{Hannah Wong}, \bibinfo{person}{Lin Yang}, \bibinfo{person}{Yu Yang}, \bibinfo{person}{Michihiro Yasunaga}, \bibinfo{person}{Kristen Ying}, \bibinfo{person}{Wojciech Zaremba}, \bibinfo{person}{Wenting Zhan}, \bibinfo{person}{Cyril Zhang}, \bibinfo{person}{Brian Zhang}, \bibinfo{person}{Eddie Zhang}, {and} \bibinfo{person}{Shengjia Zhao}.}
  \bibinfo{year}{2025}\natexlab{}.
\newblock \bibinfo{title}{gpt-oss-120b \& gpt-oss-20b {Model} {Card}}.
\newblock
\href{https://doi.org/10.48550/arXiv.2508.10925}{doi:\nolinkurl{10.48550/arXiv.2508.10925}}
\newblock
\shownote{arXiv:2508.10925 [cs]}.


\bibitem[Radford et~al\mbox{.}(2023)]%
        {10.5555/3618408.3619590}
\bibfield{author}{\bibinfo{person}{Alec Radford}, \bibinfo{person}{Jong~Wook Kim}, \bibinfo{person}{Tao Xu}, \bibinfo{person}{Greg Brockman}, \bibinfo{person}{Christine McLeavey}, {and} \bibinfo{person}{Ilya Sutskever}.} \bibinfo{year}{2023}\natexlab{}.
\newblock \showarticletitle{Robust speech recognition via large-scale weak supervision}. In \bibinfo{booktitle}{\emph{Proceedings of the 40th International Conference on Machine Learning}} (Honolulu, Hawaii, USA) \emph{(\bibinfo{series}{ICML'23})}. \bibinfo{publisher}{JMLR.org}, Article \bibinfo{articleno}{1182}, \bibinfo{numpages}{27}~pages.
\newblock


\bibitem[Rietz and Maedche(2021)]%
        {rietz_cody_2021}
\bibfield{author}{\bibinfo{person}{Tim Rietz} {and} \bibinfo{person}{Alexander Maedche}.} \bibinfo{year}{2021}\natexlab{}.
\newblock \showarticletitle{Cody: {An} {AI}-{Based} {System} to {Semi}-{Automate} {Coding} for {Qualitative} {Research}}. In \bibinfo{booktitle}{\emph{Proceedings of the 2021 {CHI} {Conference} on {Human} {Factors} in {Computing} {Systems}}}. \bibinfo{publisher}{ACM}, \bibinfo{address}{Yokohama Japan}, \bibinfo{pages}{1--14}.
\newblock
\showISBNx{978-1-4503-8096-6}
\href{https://doi.org/10.1145/3411764.3445591}{doi:\nolinkurl{10.1145/3411764.3445591}}


\bibitem[Sam and Vavekanand(2024)]%
        {sam_llama_2024}
\bibfield{author}{\bibinfo{person}{Kira Sam} {and} \bibinfo{person}{Raja Vavekanand}.} \bibinfo{year}{2024}\natexlab{}.
\newblock \showarticletitle{Llama 3.1: {An} {In}-{Depth} {Analysis} of the {Next} {Generation} {Large} {Language} {Model}}.
\newblock  (\bibinfo{year}{2024}).
\newblock
\href{https://doi.org/10.13140/RG.2.2.10628.74882}{doi:\nolinkurl{10.13140/RG.2.2.10628.74882}}
\newblock
\shownote{Publisher: Unpublished}.


\bibitem[Sinha et~al\mbox{.}(2024)]%
        {sinha_role_2024}
\bibfield{author}{\bibinfo{person}{Ravi Sinha}, \bibinfo{person}{Idris Solola}, \bibinfo{person}{Ha Nguyen}, \bibinfo{person}{Hillary Swanson}, {and} \bibinfo{person}{LuEttaMae Lawrence}.} \bibinfo{year}{2024}\natexlab{}.
\newblock \showarticletitle{The {Role} of {Generative} {AI} in {Qualitative} {Research}: {GPT}-4's {Contributions} to a {Grounded} {Theory} {Analysis}}. In \bibinfo{booktitle}{\emph{Proceedings of the {Symposium} on {Learning} {Design} and {Technology}}}. \bibinfo{publisher}{ACM}, \bibinfo{address}{Delft Netherlands}, \bibinfo{pages}{17--25}.
\newblock
\showISBNx{979-8-4007-1722-2}
\href{https://doi.org/10.1145/3663433.3663456}{doi:\nolinkurl{10.1145/3663433.3663456}}


\bibitem[Team et~al\mbox{.}(2025b)]%
        {team_glm-45_2025}
\bibfield{author}{\bibinfo{person}{GLM-4~5 Team}, \bibinfo{person}{Aohan Zeng}, \bibinfo{person}{Xin Lv}, \bibinfo{person}{Qinkai Zheng}, \bibinfo{person}{Zhenyu Hou}, \bibinfo{person}{Bin Chen}, \bibinfo{person}{Chengxing Xie}, \bibinfo{person}{Cunxiang Wang}, \bibinfo{person}{Da Yin}, \bibinfo{person}{Hao Zeng}, \bibinfo{person}{Jiajie Zhang}, \bibinfo{person}{Kedong Wang}, \bibinfo{person}{Lucen Zhong}, \bibinfo{person}{Mingdao Liu}, \bibinfo{person}{Rui Lu}, \bibinfo{person}{Shulin Cao}, \bibinfo{person}{Xiaohan Zhang}, \bibinfo{person}{Xuancheng Huang}, \bibinfo{person}{Yao Wei}, \bibinfo{person}{Yean Cheng}, \bibinfo{person}{Yifan An}, \bibinfo{person}{Yilin Niu}, \bibinfo{person}{Yuanhao Wen}, \bibinfo{person}{Yushi Bai}, \bibinfo{person}{Zhengxiao Du}, \bibinfo{person}{Zihan Wang}, \bibinfo{person}{Zilin Zhu}, \bibinfo{person}{Bohan Zhang}, \bibinfo{person}{Bosi Wen}, \bibinfo{person}{Bowen Wu}, \bibinfo{person}{Bowen Xu}, \bibinfo{person}{Can Huang}, \bibinfo{person}{Casey Zhao},
  \bibinfo{person}{Changpeng Cai}, \bibinfo{person}{Chao Yu}, \bibinfo{person}{Chen Li}, \bibinfo{person}{Chendi Ge}, \bibinfo{person}{Chenghua Huang}, \bibinfo{person}{Chenhui Zhang}, \bibinfo{person}{Chenxi Xu}, \bibinfo{person}{Chenzheng Zhu}, \bibinfo{person}{Chuang Li}, \bibinfo{person}{Congfeng Yin}, \bibinfo{person}{Daoyan Lin}, \bibinfo{person}{Dayong Yang}, \bibinfo{person}{Dazhi Jiang}, \bibinfo{person}{Ding Ai}, \bibinfo{person}{Erle Zhu}, \bibinfo{person}{Fei Wang}, \bibinfo{person}{Gengzheng Pan}, \bibinfo{person}{Guo Wang}, \bibinfo{person}{Hailong Sun}, \bibinfo{person}{Haitao Li}, \bibinfo{person}{Haiyang Li}, \bibinfo{person}{Haiyi Hu}, \bibinfo{person}{Hanyu Zhang}, \bibinfo{person}{Hao Peng}, \bibinfo{person}{Hao Tai}, \bibinfo{person}{Haoke Zhang}, \bibinfo{person}{Haoran Wang}, \bibinfo{person}{Haoyu Yang}, \bibinfo{person}{He Liu}, \bibinfo{person}{He Zhao}, \bibinfo{person}{Hongwei Liu}, \bibinfo{person}{Hongxi Yan}, \bibinfo{person}{Huan Liu}, \bibinfo{person}{Huilong Chen},
  \bibinfo{person}{Ji Li}, \bibinfo{person}{Jiajing Zhao}, \bibinfo{person}{Jiamin Ren}, \bibinfo{person}{Jian Jiao}, \bibinfo{person}{Jiani Zhao}, \bibinfo{person}{Jianyang Yan}, \bibinfo{person}{Jiaqi Wang}, \bibinfo{person}{Jiayi Gui}, \bibinfo{person}{Jiayue Zhao}, \bibinfo{person}{Jie Liu}, \bibinfo{person}{Jijie Li}, \bibinfo{person}{Jing Li}, \bibinfo{person}{Jing Lu}, \bibinfo{person}{Jingsen Wang}, \bibinfo{person}{Jingwei Yuan}, \bibinfo{person}{Jingxuan Li}, \bibinfo{person}{Jingzhao Du}, \bibinfo{person}{Jinhua Du}, \bibinfo{person}{Jinxin Liu}, \bibinfo{person}{Junkai Zhi}, \bibinfo{person}{Junli Gao}, \bibinfo{person}{Ke Wang}, \bibinfo{person}{Lekang Yang}, \bibinfo{person}{Liang Xu}, \bibinfo{person}{Lin Fan}, \bibinfo{person}{Lindong Wu}, \bibinfo{person}{Lintao Ding}, \bibinfo{person}{Lu Wang}, \bibinfo{person}{Man Zhang}, \bibinfo{person}{Minghao Li}, \bibinfo{person}{Minghuan Xu}, \bibinfo{person}{Mingming Zhao}, \bibinfo{person}{Mingshu Zhai}, \bibinfo{person}{Pengfan Du},
  \bibinfo{person}{Qian Dong}, \bibinfo{person}{Shangde Lei}, \bibinfo{person}{Shangqing Tu}, \bibinfo{person}{Shangtong Yang}, \bibinfo{person}{Shaoyou Lu}, \bibinfo{person}{Shijie Li}, \bibinfo{person}{Shuang Li}, \bibinfo{person}{Shuang-Li}, \bibinfo{person}{Shuxun Yang}, \bibinfo{person}{Sibo Yi}, \bibinfo{person}{Tianshu Yu}, \bibinfo{person}{Wei Tian}, \bibinfo{person}{Weihan Wang}, \bibinfo{person}{Wenbo Yu}, \bibinfo{person}{Weng~Lam Tam}, \bibinfo{person}{Wenjie Liang}, \bibinfo{person}{Wentao Liu}, \bibinfo{person}{Xiao Wang}, \bibinfo{person}{Xiaohan Jia}, \bibinfo{person}{Xiaotao Gu}, \bibinfo{person}{Xiaoying Ling}, \bibinfo{person}{Xin Wang}, \bibinfo{person}{Xing Fan}, \bibinfo{person}{Xingru Pan}, \bibinfo{person}{Xinyuan Zhang}, \bibinfo{person}{Xinze Zhang}, \bibinfo{person}{Xiuqing Fu}, \bibinfo{person}{Xunkai Zhang}, \bibinfo{person}{Yabo Xu}, \bibinfo{person}{Yandong Wu}, \bibinfo{person}{Yida Lu}, \bibinfo{person}{Yidong Wang}, \bibinfo{person}{Yilin Zhou}, \bibinfo{person}{Yiming Pan},
  \bibinfo{person}{Ying Zhang}, \bibinfo{person}{Yingli Wang}, \bibinfo{person}{Yingru Li}, \bibinfo{person}{Yinpei Su}, \bibinfo{person}{Yipeng Geng}, \bibinfo{person}{Yitong Zhu}, \bibinfo{person}{Yongkun Yang}, \bibinfo{person}{Yuhang Li}, \bibinfo{person}{Yuhao Wu}, \bibinfo{person}{Yujiang Li}, \bibinfo{person}{Yunan Liu}, \bibinfo{person}{Yunqing Wang}, \bibinfo{person}{Yuntao Li}, \bibinfo{person}{Yuxuan Zhang}, \bibinfo{person}{Zezhen Liu}, \bibinfo{person}{Zhen Yang}, \bibinfo{person}{Zhengda Zhou}, \bibinfo{person}{Zhongpei Qiao}, \bibinfo{person}{Zhuoer Feng}, \bibinfo{person}{Zhuorui Liu}, \bibinfo{person}{Zichen Zhang}, \bibinfo{person}{Zihan Wang}, \bibinfo{person}{Zijun Yao}, \bibinfo{person}{Zikang Wang}, \bibinfo{person}{Ziqiang Liu}, \bibinfo{person}{Ziwei Chai}, \bibinfo{person}{Zixuan Li}, \bibinfo{person}{Zuodong Zhao}, \bibinfo{person}{Wenguang Chen}, \bibinfo{person}{Jidong Zhai}, \bibinfo{person}{Bin Xu}, \bibinfo{person}{Minlie Huang}, \bibinfo{person}{Hongning Wang},
  \bibinfo{person}{Juanzi Li}, \bibinfo{person}{Yuxiao Dong}, {and} \bibinfo{person}{Jie Tang}.} \bibinfo{year}{2025}\natexlab{b}.
\newblock \bibinfo{title}{{GLM}-4.5: {Agentic}, {Reasoning}, and {Coding} ({ARC}) {Foundation} {Models}}.
\newblock
\href{https://doi.org/10.48550/arXiv.2508.06471}{doi:\nolinkurl{10.48550/arXiv.2508.06471}}
\newblock
\shownote{arXiv:2508.06471 [cs]}.


\bibitem[Team et~al\mbox{.}(2025a)]%
        {team_kimi_2025}
\bibfield{author}{\bibinfo{person}{Kimi Team}, \bibinfo{person}{Yifan Bai}, \bibinfo{person}{Yiping Bao}, \bibinfo{person}{Guanduo Chen}, \bibinfo{person}{Jiahao Chen}, \bibinfo{person}{Ningxin Chen}, \bibinfo{person}{Ruijue Chen}, \bibinfo{person}{Yanru Chen}, \bibinfo{person}{Yuankun Chen}, \bibinfo{person}{Yutian Chen}, \bibinfo{person}{Zhuofu Chen}, \bibinfo{person}{Jialei Cui}, \bibinfo{person}{Hao Ding}, \bibinfo{person}{Mengnan Dong}, \bibinfo{person}{Angang Du}, \bibinfo{person}{Chenzhuang Du}, \bibinfo{person}{Dikang Du}, \bibinfo{person}{Yulun Du}, \bibinfo{person}{Yu Fan}, \bibinfo{person}{Yichen Feng}, \bibinfo{person}{Kelin Fu}, \bibinfo{person}{Bofei Gao}, \bibinfo{person}{Hongcheng Gao}, \bibinfo{person}{Peizhong Gao}, \bibinfo{person}{Tong Gao}, \bibinfo{person}{Xinran Gu}, \bibinfo{person}{Longyu Guan}, \bibinfo{person}{Haiqing Guo}, \bibinfo{person}{Jianhang Guo}, \bibinfo{person}{Hao Hu}, \bibinfo{person}{Xiaoru Hao}, \bibinfo{person}{Tianhong He}, \bibinfo{person}{Weiran He},
  \bibinfo{person}{Wenyang He}, \bibinfo{person}{Chao Hong}, \bibinfo{person}{Yangyang Hu}, \bibinfo{person}{Zhenxing Hu}, \bibinfo{person}{Weixiao Huang}, \bibinfo{person}{Zhiqi Huang}, \bibinfo{person}{Zihao Huang}, \bibinfo{person}{Tao Jiang}, \bibinfo{person}{Zhejun Jiang}, \bibinfo{person}{Xinyi Jin}, \bibinfo{person}{Yongsheng Kang}, \bibinfo{person}{Guokun Lai}, \bibinfo{person}{Cheng Li}, \bibinfo{person}{Fang Li}, \bibinfo{person}{Haoyang Li}, \bibinfo{person}{Ming Li}, \bibinfo{person}{Wentao Li}, \bibinfo{person}{Yanhao Li}, \bibinfo{person}{Yiwei Li}, \bibinfo{person}{Zhaowei Li}, \bibinfo{person}{Zheming Li}, \bibinfo{person}{Hongzhan Lin}, \bibinfo{person}{Xiaohan Lin}, \bibinfo{person}{Zongyu Lin}, \bibinfo{person}{Chengyin Liu}, \bibinfo{person}{Chenyu Liu}, \bibinfo{person}{Hongzhang Liu}, \bibinfo{person}{Jingyuan Liu}, \bibinfo{person}{Junqi Liu}, \bibinfo{person}{Liang Liu}, \bibinfo{person}{Shaowei Liu}, \bibinfo{person}{T.~Y. Liu}, \bibinfo{person}{Tianwei Liu}, \bibinfo{person}{Weizhou
  Liu}, \bibinfo{person}{Yangyang Liu}, \bibinfo{person}{Yibo Liu}, \bibinfo{person}{Yiping Liu}, \bibinfo{person}{Yue Liu}, \bibinfo{person}{Zhengying Liu}, \bibinfo{person}{Enzhe Lu}, \bibinfo{person}{Lijun Lu}, \bibinfo{person}{Shengling Ma}, \bibinfo{person}{Xinyu Ma}, \bibinfo{person}{Yingwei Ma}, \bibinfo{person}{Shaoguang Mao}, \bibinfo{person}{Jie Mei}, \bibinfo{person}{Xin Men}, \bibinfo{person}{Yibo Miao}, \bibinfo{person}{Siyuan Pan}, \bibinfo{person}{Yebo Peng}, \bibinfo{person}{Ruoyu Qin}, \bibinfo{person}{Bowen Qu}, \bibinfo{person}{Zeyu Shang}, \bibinfo{person}{Lidong Shi}, \bibinfo{person}{Shengyuan Shi}, \bibinfo{person}{Feifan Song}, \bibinfo{person}{Jianlin Su}, \bibinfo{person}{Zhengyuan Su}, \bibinfo{person}{Xinjie Sun}, \bibinfo{person}{Flood Sung}, \bibinfo{person}{Heyi Tang}, \bibinfo{person}{Jiawen Tao}, \bibinfo{person}{Qifeng Teng}, \bibinfo{person}{Chensi Wang}, \bibinfo{person}{Dinglu Wang}, \bibinfo{person}{Feng Wang}, \bibinfo{person}{Haiming Wang}, \bibinfo{person}{Jianzhou
  Wang}, \bibinfo{person}{Jiaxing Wang}, \bibinfo{person}{Jinhong Wang}, \bibinfo{person}{Shengjie Wang}, \bibinfo{person}{Shuyi Wang}, \bibinfo{person}{Yao Wang}, \bibinfo{person}{Yejie Wang}, \bibinfo{person}{Yiqin Wang}, \bibinfo{person}{Yuxin Wang}, \bibinfo{person}{Yuzhi Wang}, \bibinfo{person}{Zhaoji Wang}, \bibinfo{person}{Zhengtao Wang}, \bibinfo{person}{Zhexu Wang}, \bibinfo{person}{Chu Wei}, \bibinfo{person}{Qianqian Wei}, \bibinfo{person}{Wenhao Wu}, \bibinfo{person}{Xingzhe Wu}, \bibinfo{person}{Yuxin Wu}, \bibinfo{person}{Chenjun Xiao}, \bibinfo{person}{Xiaotong Xie}, \bibinfo{person}{Weimin Xiong}, \bibinfo{person}{Boyu Xu}, \bibinfo{person}{Jing Xu}, \bibinfo{person}{Jinjing Xu}, \bibinfo{person}{L.~H. Xu}, \bibinfo{person}{Lin Xu}, \bibinfo{person}{Suting Xu}, \bibinfo{person}{Weixin Xu}, \bibinfo{person}{Xinran Xu}, \bibinfo{person}{Yangchuan Xu}, \bibinfo{person}{Ziyao Xu}, \bibinfo{person}{Junjie Yan}, \bibinfo{person}{Yuzi Yan}, \bibinfo{person}{Xiaofei Yang}, \bibinfo{person}{Ying Yang},
  \bibinfo{person}{Zhen Yang}, \bibinfo{person}{Zhilin Yang}, \bibinfo{person}{Zonghan Yang}, \bibinfo{person}{Haotian Yao}, \bibinfo{person}{Xingcheng Yao}, \bibinfo{person}{Wenjie Ye}, \bibinfo{person}{Zhuorui Ye}, \bibinfo{person}{Bohong Yin}, \bibinfo{person}{Longhui Yu}, \bibinfo{person}{Enming Yuan}, \bibinfo{person}{Hongbang Yuan}, \bibinfo{person}{Mengjie Yuan}, \bibinfo{person}{Haobing Zhan}, \bibinfo{person}{Dehao Zhang}, \bibinfo{person}{Hao Zhang}, \bibinfo{person}{Wanlu Zhang}, \bibinfo{person}{Xiaobin Zhang}, \bibinfo{person}{Yangkun Zhang}, \bibinfo{person}{Yizhi Zhang}, \bibinfo{person}{Yongting Zhang}, \bibinfo{person}{Yu Zhang}, \bibinfo{person}{Yutao Zhang}, \bibinfo{person}{Yutong Zhang}, \bibinfo{person}{Zheng Zhang}, \bibinfo{person}{Haotian Zhao}, \bibinfo{person}{Yikai Zhao}, \bibinfo{person}{Huabin Zheng}, \bibinfo{person}{Shaojie Zheng}, \bibinfo{person}{Jianren Zhou}, \bibinfo{person}{Xinyu Zhou}, \bibinfo{person}{Zaida Zhou}, \bibinfo{person}{Zhen Zhu}, \bibinfo{person}{Weiyu
  Zhuang}, {and} \bibinfo{person}{Xinxing Zu}.} \bibinfo{year}{2025}\natexlab{a}.
\newblock \bibinfo{title}{Kimi {K2}: {Open} {Agentic} {Intelligence}}.
\newblock
\href{https://doi.org/10.48550/arXiv.2507.20534}{doi:\nolinkurl{10.48550/arXiv.2507.20534}}
\newblock
\shownote{arXiv:2507.20534 [cs]}.


\bibitem[Xiao et~al\mbox{.}(2023)]%
        {xiao_supporting_2023}
\bibfield{author}{\bibinfo{person}{Ziang Xiao}, \bibinfo{person}{Xingdi Yuan}, \bibinfo{person}{Q.~Vera Liao}, \bibinfo{person}{Rania Abdelghani}, {and} \bibinfo{person}{Pierre-Yves Oudeyer}.} \bibinfo{year}{2023}\natexlab{}.
\newblock \showarticletitle{Supporting {Qualitative} {Analysis} with {Large} {Language} {Models}: {Combining} {Codebook} with {GPT}-3 for {Deductive} {Coding}}. In \bibinfo{booktitle}{\emph{28th {International} {Conference} on {Intelligent} {User} {Interfaces}}}. \bibinfo{publisher}{ACM}, \bibinfo{address}{Sydney NSW Australia}, \bibinfo{pages}{75--78}.
\newblock
\showISBNx{979-8-4007-0107-8}
\href{https://doi.org/10.1145/3581754.3584136}{doi:\nolinkurl{10.1145/3581754.3584136}}


\bibitem[Yan et~al\mbox{.}(2024)]%
        {yan_human-ai_2024}
\bibfield{author}{\bibinfo{person}{Lixiang Yan}, \bibinfo{person}{Vanessa Echeverria}, \bibinfo{person}{Gloria~Milena Fernandez-Nieto}, \bibinfo{person}{Yueqiao Jin}, \bibinfo{person}{Zachari Swiecki}, \bibinfo{person}{Linxuan Zhao}, \bibinfo{person}{Dragan Gašević}, {and} \bibinfo{person}{Roberto Martinez-Maldonado}.} \bibinfo{year}{2024}\natexlab{}.
\newblock \showarticletitle{Human-{AI} {Collaboration} in {Thematic} {Analysis} using {ChatGPT}: {A} {User} {Study} and {Design} {Recommendations}}. In \bibinfo{booktitle}{\emph{Extended {Abstracts} of the {CHI} {Conference} on {Human} {Factors} in {Computing} {Systems}}}. \bibinfo{publisher}{ACM}, \bibinfo{address}{Honolulu HI USA}, \bibinfo{pages}{1--7}.
\newblock
\showISBNx{979-8-4007-0331-7}
\href{https://doi.org/10.1145/3613905.3650732}{doi:\nolinkurl{10.1145/3613905.3650732}}


\bibitem[Yang et~al\mbox{.}(2025)]%
        {yang_qwen3_2025}
\bibfield{author}{\bibinfo{person}{An Yang}, \bibinfo{person}{Anfeng Li}, \bibinfo{person}{Baosong Yang}, \bibinfo{person}{Beichen Zhang}, \bibinfo{person}{Binyuan Hui}, \bibinfo{person}{Bo Zheng}, \bibinfo{person}{Bowen Yu}, \bibinfo{person}{Chang Gao}, \bibinfo{person}{Chengen Huang}, \bibinfo{person}{Chenxu Lv}, \bibinfo{person}{Chujie Zheng}, \bibinfo{person}{Dayiheng Liu}, \bibinfo{person}{Fan Zhou}, \bibinfo{person}{Fei Huang}, \bibinfo{person}{Feng Hu}, \bibinfo{person}{Hao Ge}, \bibinfo{person}{Haoran Wei}, \bibinfo{person}{Huan Lin}, \bibinfo{person}{Jialong Tang}, \bibinfo{person}{Jian Yang}, \bibinfo{person}{Jianhong Tu}, \bibinfo{person}{Jianwei Zhang}, \bibinfo{person}{Jianxin Yang}, \bibinfo{person}{Jiaxi Yang}, \bibinfo{person}{Jing Zhou}, \bibinfo{person}{Jingren Zhou}, \bibinfo{person}{Junyang Lin}, \bibinfo{person}{Kai Dang}, \bibinfo{person}{Keqin Bao}, \bibinfo{person}{Kexin Yang}, \bibinfo{person}{Le Yu}, \bibinfo{person}{Lianghao Deng}, \bibinfo{person}{Mei Li}, \bibinfo{person}{Mingfeng
  Xue}, \bibinfo{person}{Mingze Li}, \bibinfo{person}{Pei Zhang}, \bibinfo{person}{Peng Wang}, \bibinfo{person}{Qin Zhu}, \bibinfo{person}{Rui Men}, \bibinfo{person}{Ruize Gao}, \bibinfo{person}{Shixuan Liu}, \bibinfo{person}{Shuang Luo}, \bibinfo{person}{Tianhao Li}, \bibinfo{person}{Tianyi Tang}, \bibinfo{person}{Wenbiao Yin}, \bibinfo{person}{Xingzhang Ren}, \bibinfo{person}{Xinyu Wang}, \bibinfo{person}{Xinyu Zhang}, \bibinfo{person}{Xuancheng Ren}, \bibinfo{person}{Yang Fan}, \bibinfo{person}{Yang Su}, \bibinfo{person}{Yichang Zhang}, \bibinfo{person}{Yinger Zhang}, \bibinfo{person}{Yu Wan}, \bibinfo{person}{Yuqiong Liu}, \bibinfo{person}{Zekun Wang}, \bibinfo{person}{Zeyu Cui}, \bibinfo{person}{Zhenru Zhang}, \bibinfo{person}{Zhipeng Zhou}, {and} \bibinfo{person}{Zihan Qiu}.} \bibinfo{year}{2025}\natexlab{}.
\newblock \bibinfo{title}{Qwen3 {Technical} {Report}}.
\newblock
\href{https://doi.org/10.48550/arXiv.2505.09388}{doi:\nolinkurl{10.48550/arXiv.2505.09388}}
\newblock
\shownote{arXiv:2505.09388 [cs]}.


\bibitem[Yao et~al\mbox{.}(2024)]%
        {yao_survey_2024}
\bibfield{author}{\bibinfo{person}{Yifan Yao}, \bibinfo{person}{Jinhao Duan}, \bibinfo{person}{Kaidi Xu}, \bibinfo{person}{Yuanfang Cai}, \bibinfo{person}{Zhibo Sun}, {and} \bibinfo{person}{Yue Zhang}.} \bibinfo{year}{2024}\natexlab{}.
\newblock \showarticletitle{A survey on large language model ({LLM}) security and privacy: {The} {Good}, {The} {Bad}, and {The} {Ugly}}.
\newblock \bibinfo{journal}{\emph{High-Confidence Computing}} \bibinfo{volume}{4}, \bibinfo{number}{2} (\bibinfo{date}{June} \bibinfo{year}{2024}), \bibinfo{pages}{100211}.
\newblock
\showISSN{26672952}
\href{https://doi.org/10.1016/j.hcc.2024.100211}{doi:\nolinkurl{10.1016/j.hcc.2024.100211}}


\bibitem[Yin(2018)]%
        {1970586434844156305}
\bibfield{author}{\bibinfo{person}{Robert~K. Yin}.} \bibinfo{year}{2018}\natexlab{}.
\newblock \bibinfo{booktitle}{\emph{Case study research and applications : design and methods} (\bibinfo{edition}{6th ed} ed.)}.
\newblock \bibinfo{publisher}{Sage}.
\newblock
\urldef\tempurl%
\url{https://cir.nii.ac.jp/crid/1970586434844156305}
\showURL{%
\tempurl}


\end{thebibliography}

\end{document}